\documentclass[iop]{emulateapj}

\usepackage{graphicx}

\def\kms{km~s$^{-1}$} 
\usepackage{amsmath}
\slugcomment{Accepted to The Astrophysical Journal, October 26, 2015}

\shorttitle{The Remarkable Metamorphosis of SN\,2014C}
\shortauthors{Milisavljevic et al.}

\begin{document}

\def\cfa{1}
\def\uaz{2}
\def\ua{3}
\def\einstein{4}
\def\hart{5}
\def\york{6}
\def\uo{7}
\def\su{8}
\def\dartmouth{9}
\def\aia{10}

\title{Metamorphosis of SN\,2014C: Delayed interaction between a
  hydrogen poor core-collapse supernova and a nearby circumstellar
  shell}

\author{D.~Milisavljevic\altaffilmark{\cfa}, 
R.~Margutti\altaffilmark{\cfa},
A.~Kamble\altaffilmark{\cfa},
D.~J.~Patnaude\altaffilmark{\cfa},
J.~C.~Raymond\altaffilmark{\cfa},
J.~J.~Eldridge\altaffilmark{\uaz}, 
W.~Fong\altaffilmark{\ua,\einstein},
M.~Bietenholz\altaffilmark{\hart,\york},
P.~Challis\altaffilmark{\cfa},
R.~Chornock\altaffilmark{\uo},
M.~R.~Drout\altaffilmark{\cfa},
C.~Fransson\altaffilmark{\su},
R.~A.~Fesen\altaffilmark{\dartmouth},\\
J.~E.~Grindlay\altaffilmark{\cfa},
R.~P.~Kirshner\altaffilmark{\cfa},
R.~Lunnan\altaffilmark{\cfa},
J.~Mackey\altaffilmark{\aia},
G.~F.~Miller\altaffilmark{\cfa},\\
J.~T.~Parrent\altaffilmark{\cfa},
N.~E.~Sanders\altaffilmark{\cfa},
A.~M.~Soderberg\altaffilmark{\cfa},
B.~A.~Zauderer\altaffilmark{\cfa}
} 

\altaffiltext{\cfa}{Harvard-Smithsonian Center for Astrophysics, 60
  Garden Street, Cambridge, MA, 02138. \\ Electronic address:
  dmilisav@cfa.harvard.edu} 
\altaffiltext{\uaz}{Department of Physics,
  University of Auckland, Private Bag 92019, Auckland, New Zealand}
\altaffiltext{\ua}{Steward Observatory, University of Arizona, 933
  N. Cherry Ave, Tucson, AZ 85721} 
\altaffiltext{\einstein}{NASA
  Einstein Fellow} 
\altaffiltext{\hart}{Hartebeesthoek Radio
  Observatory, PO Box 443, Krugersdorp 1740, South Africa}
\altaffiltext{\york}{Department of Physics and Astronomy, York
  University, Toronto, ON M3J 1P3, Canada}
\altaffiltext{\uo}{Astrophysical Institute, Department of Physics and
  Astronomy, 251B Clippinger Lab, Ohio University, Athens, OH 45701,
  USA} 
\altaffiltext{\su}{Oskar Klein Centre, Department of Astronomy,
  Stockholm University, AlbaNova, SE–106 91 Stockholm, Sweden}
\altaffiltext{\dartmouth}{Department of Physics \& Astronomy,
  Dartmouth College, 6127 Wilder Lab, Hanover, NH 03755, USA}
\altaffiltext{\aia}{Argelander-Institut f\''ur Astronomie, Auf dem Hügel
  71, 53121 Bonn, Germany}

\begin{abstract}

  We present optical observations of supernova SN\,2014C, which
  underwent an unprecedented slow metamorphosis from H-poor type Ib to
  H-rich type IIn over the course of one year. The observed
  spectroscopic evolution is consistent with the supernova having
  exploded in a cavity before encountering a massive shell of the
  progenitor star's stripped hydrogen envelope. Possible origins for
  the circumstellar shell include a brief Wolf-Rayet fast wind phase
  that overtook a slower red supergiant wind, eruptive ejection, or
  confinement of circumstellar material by external influences of
  neighboring stars. An extended high velocity H$\alpha$ absorption
  feature seen in near-maximum light spectra implies that the
  progenitor star was not completely stripped of hydrogen at the time
  of core collapse. Archival pre-explosion Subaru Telescope
  Suprime-Cam and Hubble Space Telescope Wide Field Planetary Camera~2
  images of the region obtained in 2009 show a coincident source that
  is most likely a compact massive star cluster in NGC\,7331 that
  hosted the progenitor system. By comparing the emission properties
  of the source with stellar population models that incorporate
  interacting binary stars we estimate the age of the host cluster to
  be $30 - 300$ Myr, and favor ages closer to $30$ Myr in light of
  relatively strong H$\alpha$ emission.  SN\,2014C is the
  best-observed member of a class of core-collapse supernovae that
  fill the gap between events that interact strongly with dense,
  nearby environments immediately after explosion and those that never
  show signs of interaction. Better understanding of the frequency and
  nature of this intermediate population can contribute valuable
  information about the poorly understood final stages of stellar
  evolution.

\end{abstract}


\section{Introduction}
\label{sec:Intro}

Recent observations have convincingly demonstrated that massive stars
can experience an eruptive mass loss episode $\sim 1$ yr before core
collapse. Such episodes have been confirmed in H-rich type IIn
\citep{Ofek13,Smith14,Ofek14} and H-poor, He-rich type Ibn
\citep{Pastorello07,Foley07} supernovae.  Although the brief
timescales between eruption and supernova explosion have been
anticipated in special cases of very massive stars
\citep{Woosley07,Quataert12}, a growing number of systems are being
discovered that fall outside most theoretical regimes and that
challenge many long held notions of stellar evolution
\citep{Stritzinger12,Pastorello13,Mauerhan13,Margutti14,SmithArnett14,Moriya15}.

The most favored candidate progenitors of type IIn and Ibn supernovae
are luminous blue variable (LBV) stars
\citep{Kotak06,Pastorello07,Gal-Yam09-05gl}, and stars leaving the LBV
phase and entering a Wolf-Rayet (W-R) phase
\citep{Smith12,Pastorello15} that may be exceptionally brief ($\la
10^3$ yr; \citealt{Dwarkadas11}).  LBV stars are prime suspects
because they undergo recurrent mass-loss episodes that eject large
portions of their hydrogen envelope ($\ga 1\;M_{\odot}$;
\citealt{Smith06}). However, traditional stellar evolutionary theory
has predicted LBV stars to be a transitional phase lasting $10^4$ to
$10^5$ yr before evolving to a compact and hydrogen-poor W-R star
lasting a few $\times 10^5$ yr
\citep{Meynet94,Langer94,Maeder00}. This predicted timescale is
discrepant with the $\sim 1$ yr timescale between eruption and core
collapse observed in at least some type IIn and Ibn systems
\citep{Pastorello07,Pastorello13,Smith14}.

\begin{figure*}[htp!]
\centering 

\includegraphics[width=\linewidth]{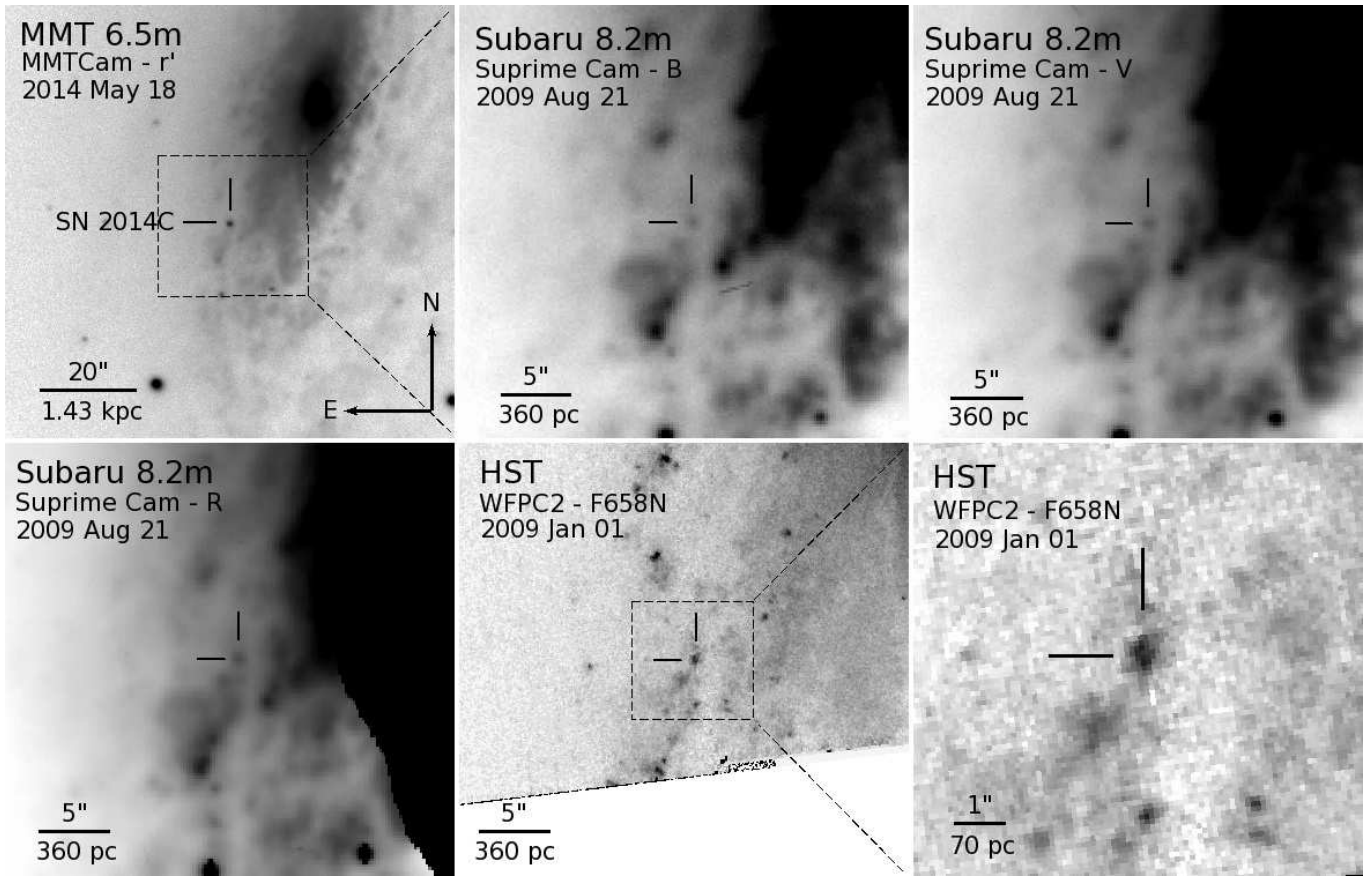}

\caption{Finding chart of SN\,2014C and pre-explosion images. Top
  left: MMT 6.5m telescope $r'$-band image of the region around
  SN\,2014C (marked) and its host galaxy NGC\,7331 obtained 2014 May
  18 with the MMTCam instrument. Top middle and right and
  bottom left: Archival Subaru 8.3\,m telescope $BVR$-band images
  obtained 2009 August 21 with the Suprime-Cam instrument. A visible
  source coincident with the location of the supernova is seen in all
  three filters. The galaxy nucleus in the $R$-band image is saturated
  and masks emission west of the supernova. Bottom middle and right:
  HST/WFPC2 F658N images. Dashed boxes outline enlarged
  regions shown in the adjoining panels. }

\label{fig:images}
\end{figure*}

The extent to which eruptive and/or accelerated mass loss may directly
precede core collapse in a wider range of supernovae is poorly
constrained, particularly in type IIb, Ib, and Ic supernovae (SN Ibc)
where the progenitor star has been significantly stripped of its
hydrogen envelope. Some insight into this issue was provided by
optical \citep{Maeda15} and radio+X-ray observations \citep{Kamble15}
of the type IIb SN\,2013df. Both studies concluded that the progenitor
star experienced enhanced mass loss of $(3-8) \times 10^{-5}\;
M_{\odot}\;\rm yr^{-1}$ (for wind velocity of 10 \kms) in the final
centuries leading up to the supernova explosion. Intriguingly, not all
type IIb supernovae exhibit the same mass loss enhancement (e.g.,
SN\,2011dh; \citealt{Krauss12,Maeda14,deWitt15}). The underlying physical
reasons for this dichotomy are presently unclear, but may be related
to the progenitor star size and various channels of binary interaction
\citep{Maeda15}.

W-R stars have long been suggested as an obvious progenitor of SN Ibc
because they are deficient in hydrogen \citep{Gaskell86}. However,
only progenitor stars with much cooler atmospheres than those of W-R
stars have been unambiguously detected at the explosion sites of type
IIb supernovae \citep{Aldering94,Maund11,VanDyk14}, and no secure
direct identification has yet been made of a type Ib or Ic progenitor
system \citep{vanDyk03,Smartt09,Eldridge13}. A possible exception is
the type Ib iPTF13bvn
\citep{Cao13,Groh13,Bersten14,Eldridge15,Fremling14}.

The observed number of W-R stars is probably insufficient to account
for all SN Ibc, and low mass He stars in binary systems are more
likely to be the dominant progenitor channel
\citep{Podsiadlowski92,Wellstein99,Smartt09,Smith11,Claeys11,Langer12,Eldridge13,Dessart15,Eldridge15}. However,
because W-R stars are difficult to detect in broadband images and
large populations may presently be unaccounted for
\citep{Shara13,Massey14,Massey15}, W-R stars may still represent a
significant fraction of SN Ibc.

\begin{figure*}[htp!]
\centering 

\includegraphics[width=0.95\linewidth]{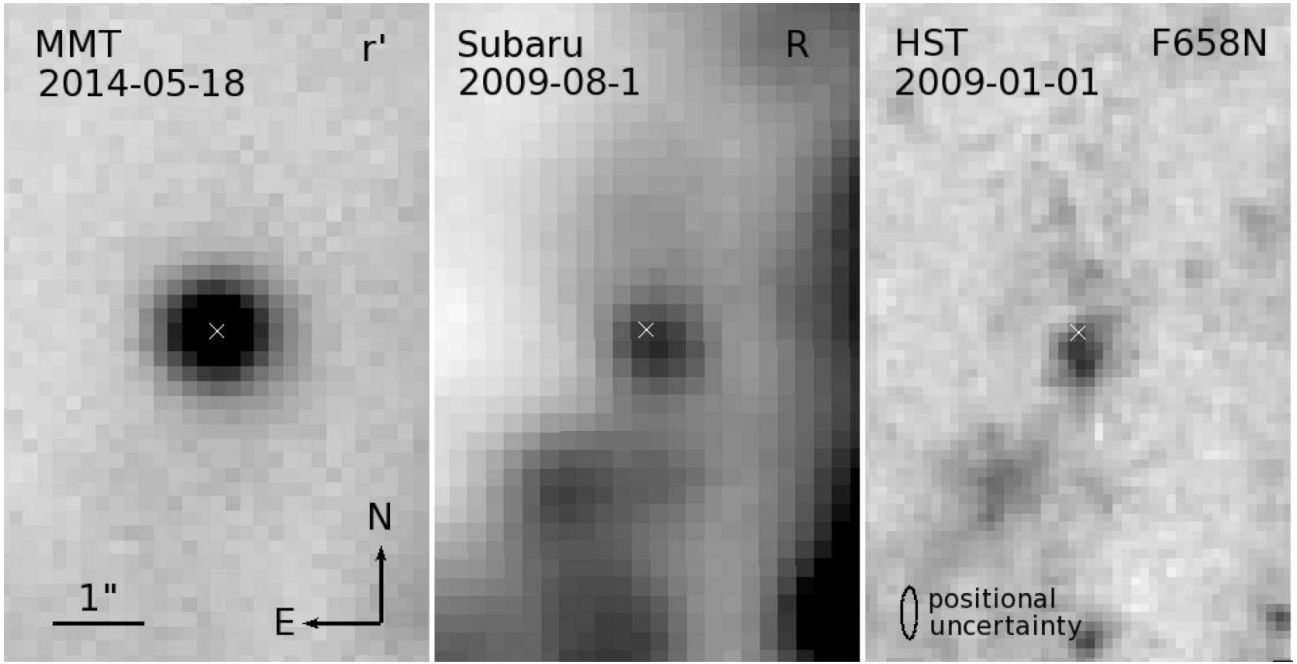}

\caption{Enlarged images of SN\,2014C and the coincident pre-explosion
  emission source. In all panels ``X'' marks the location of the
  best-fit centroid to the PSF of SN\,2014C as determined from the
  MMTCam image. A $0\farcs 24$ offset is observed between the
  supernova centroid and the peak in emission of the coincident source
  observed in the Subaru images. The ellipse in the lower left corner
  of the bottom right panel is the 3$\sigma$ error uncertainty in the
  alignment between the MMT and HST images.}

\label{fig:enlarge}
\end{figure*}

In this paper we present and analyze optical observations of a
supernova that underwent a remarkable metamorphosis consistent with a
delayed interaction between the core-collapse explosion and a massive
circumstellar shell formed from the progenitor star's partially
stripped hydrogen envelope. In Section \ref{sec:Observations}, we
present multi-epoch spectra of SN\,2014C that follow its slow
spectroscopic transformation from normal type Ib to strongly
interacting type IIn, as well as archival pre-explosion images that
show a luminous source coincident with the supernova. In Section
\ref{sec:Discussion}, we discuss properties of the partially stripped
progenitor star and its host environment, as well as possible physical
mechanisms behind the formation of the circumstellar shell. Finally,
in Section~\ref{sec:Conclusions}, we summarize our conclusions and
outline how future work on late-interacting SN Ibc systems like
SN\,2014C can contribute unique information about poorly understood
stages of stellar evolution and mass loss that immediately precede
core collapse.

\section{Observations and Results}
\label{sec:Observations}

SN 2014C was discovered at coordinates $\alpha = 22^{\rm h}37^{\rm
  m}05\fs60$ and $\delta = +34\degr24'31\farcs9$ (J2000.0) in the
nearby spiral galaxy NGC 7331 on 2014 January 5.1 UT by the Lick
Observatory Supernova Search \citep{Kim14}. NGC\,7331 is known to have
hosted two previous supernovae: SN\,1959D \citep{Humason59} and
SN\,2013bu \citep{Itagaki13}. In Figure~\ref{fig:images}, a finding
chart for SN 2014C and its immediate environment made from an
$r'$-band image obtained with the 6.5\,m MMT telescope and the MMTCam
instrument\footnote{http://www.cfa.harvard.edu/mmti/wfs.html} is
shown.

Shortly after the supernova's discovery, optical spectra obtained by
J.\ Zhang \& X.\ Wang with the 2.4m telescope of Yunnan Observatories
and L.\ Tartaglia et al.\ with the Asiago 1.82m Copernico Telescope
led to the classification of the supernova as a hydrogen-poor type Ib
near maximum light. It was at this point that our group initiated a
multi-wavelength (radio-to-X-ray) observing
campaign. Optical/ultraviolet light curves and X-ray data are
presented in Margutti et al.\ (2015), and radio data are presented in
Kamble et al.\ (2015). Results from a search for pre-explosion images
and optical spectroscopy during the first year of monitoring are
presented here. Throughout this paper we adopt the Cepheid distance of
$14.7 \pm 0.6$ Mpc to the host galaxy NGC\,7331 \citep{Freedman01},
and uncertainties are quoted at the $1\sigma$ confidence level (unless
otherwise noted).

\subsection{Pre-Explosion Images}
\label{sec:pre-explosion}

NGC\,7331 is an early type spiral galaxy that has been studied closely
from X-ray to radio wavelengths (see, e.g.,
\citealt{Thilker07}). Consequently, extensive archival data exist that
cover the region of SN\,2014C prior to explosion. We retrieved optical
images originally obtained on 2009 August 21 with the 8.3\,m Subaru
Telescope and the Suprime-Cam instrument \citep{Miyazaki02}, via the
SMOKA Science Archive\footnote{http://smoka.nao.ac.jp/}. $B$, $V$, and
$R$ filters were used with exposure lengths of 200, 300, and 560\,s,
respectively. Images were bias-corrected and flat-fielded following
standard procedures with the IRAF software\footnote{The Image
  Reduction and Analysis Facility is distributed by the National
  Optical Astronomy Observatories, which are operated by the
  Association of Universities for Research in Astronomy, Inc., under
  cooperative agreement with the National Science Foundation.}, and
the absolute positions were obtained using the \texttt{IMWCS}
software\footnote{http://tdc-www.harvard.edu/wcstools/} and the US
Naval Observatory B-1.0 catalog \citep{Monet03}.

We also retrieved archival Hubble Space Telescope (HST) images
covering the location of SN\,2014C from the Mikulski Archive for Space
Telescopes. The images were obtained with the Wide Field Planetary
Camera 2 (WFPC2) and the F658N filter ($\lambda_{\rm C} = 6591$ \AA;
$\delta\lambda$ = 29 \AA) on 2009 January 1 with a total integration
time of $3\times600$\,s under program 11966 (PI: Regan). We used the
AstroDrizzle package of the DrizzlePac 2.0
software\footnote{http://drizzlepac.stsci.edu/} to remove geometric
distortion, correct for sky background variations, flag cosmic-rays,
and drizzle the individual frames together.

\begin{deluxetable*}{lccccccc}
\centering
\tablecaption{Log of spectroscopic observations}
\tablecolumns{8}
\tablewidth{0pt}
\tablehead{\colhead{Date}      &
           \colhead{MJD}    &
           \colhead{Phase}   &
           \colhead{Telescope}   &
           \colhead{Instrument}  &
           \colhead{Grating}    &
           \colhead{Range}   &
           \colhead{Resolution}  \\
           \colhead{} &
           \colhead{} &
           \colhead{(days)} &
           \colhead{}  &
           \colhead{} &
           \colhead{}   &
           \colhead{(\AA)} &
            \colhead{(\AA)}
}
\startdata
2014 Jan 09   & 56666.08 & -4   & FLWO&FAST	      & 300  & $3500-7400$ &7.0\\
2014 May 06 & 56783.43 & 113  & MMT&Blue Channel & 300 & $3300-8500$ &7.0\\
2014 Oct 22  & 56953.28 & 282  & LBT&MODS             & Dual &$5000-10000$&3.0 \\
2015 Jan 21   & 57043.08 & 373  & MMT&Blue Channel & 300 & $3500-8500$&7.0 \\
2015 Apr 25  & 57137.47 & 467  & MMT&Blue Channel &1200&$4270-5300$&1.5 
\enddata

\tablenotetext{*}{Phase is with respect to estimated $V$-band maximum
  on 2014 January 13 (MJD 56670) reported in Margutti et al.\ (2015).}

\label{tab:speclog}
\end{deluxetable*}

A visible source is seen in close proximity with SN\,2014C in all
three filters of the Subaru images (Figure~\ref{fig:images}), as well
as in the HST F658N image. The source in the Subaru images is
unresolved, with a point spread function (PSF) that we fit with a
Gaussian having a full width at half maximum (FWHM) comparable to that
observed in nearby stars ($\approx 0\farcs 8$; average of five
stars). The source seen in the HST image is resolved, non-uniform and
extended in distribution. It is slightly elongated in the north-south
direction and is contained within an approximate effective radius of
$\approx 0 \farcs 24$.

We used the \texttt{GEOMAP} task of IRAF to determine a spatial
transformation function between the MMT, Subaru, and HST images, and
the \texttt{GEOTRAN} task to apply the
transformation. Figure~\ref{fig:enlarge} shows the location of the
pre-explosion sources with respect to SN\,2014C as observed in the MMT
image. A small offset of $0\farcs 24 \pm 0.05$ exists between the
centroid of SN 2014C observed in the MMT image and the emission peak
of the source observed in the Subaru image. A small offset is also
seen between the location of SN 2014C and the approximate center of
the extended region observed in the HST F658N image.

We performed PSF photometry on the Subaru images using the
\texttt{DAOPHOT} package in IRAF in order to constrain emission
properties of the potential progenitor system. The images were
calibrated via relative photometry using 10 stars in Sloan Digital Sky
Survey (SDSS) images that cover the field of view. Photometric
transformations were made from \citet{Jordi06} to put SDSS photometry
into the $BVRI$ system. The apparent magnitudes of the coincident
source are $m_B = 22.18 \pm 0.13$, $m_V = 21.13 \pm 0.09$, and $m_R =
20.28 \pm 0.06$.  The region surrounding the source has considerable
galaxy light and the reported apparent magnitudes may overestimate the
brightness. The uncertainties reflect the error in PSF fitting and do
not include possible error due to contamination from galaxy light.

We measured the flux contained within a $0\farcs 5$ aperture centered
on the coincident emission source observed in the HST/WFPC2 F658N
image. The sum count rate $1.00 \pm 0.01$ counts s$^{-1}$ was
multiplied by the modified PHOTFLAM parameter ($9.87 \times 10^{-17}\;
\rm erg\,s^{-1}\,cm^{-2}$\,\AA$^{-1}$) and the effective bandpass of
the filter given by the RECTW parameter (39.232 \AA). The integrated
flux, which is a sum of narrow H$\alpha$, [\ion{N}{2}]
$\lambda\lambda$6548, 6583, and continuum emission, is $(3.87 \pm
0.04) \times 10^{-15}\;\rm erg\,s^{-1}\,cm^{-2}$. By estimating the
contribution from continuum emission from the spectral energy
distribution (SED) of the Subaru photometry, and assuming the
contribution from the [\ion{N}{2}] lines to be $\approx 0.13$ per cent
of the continuum subtracted flux, we derive an observed H$\alpha$ flux
of $2.9 \times 10^{-15}\;\rm erg\,s^{-1}\,cm^{-2}$, and an unabsorbed
luminosity of $4.3\times 10^{38}\;\rm erg\,s^{-1}$ after correcting
for $E(B-V)_{\rm total}$ (cf.\ Section~\ref{sec:spectra}).

\begin{figure*}[htp!]
\centering
\includegraphics[width=0.80\linewidth]{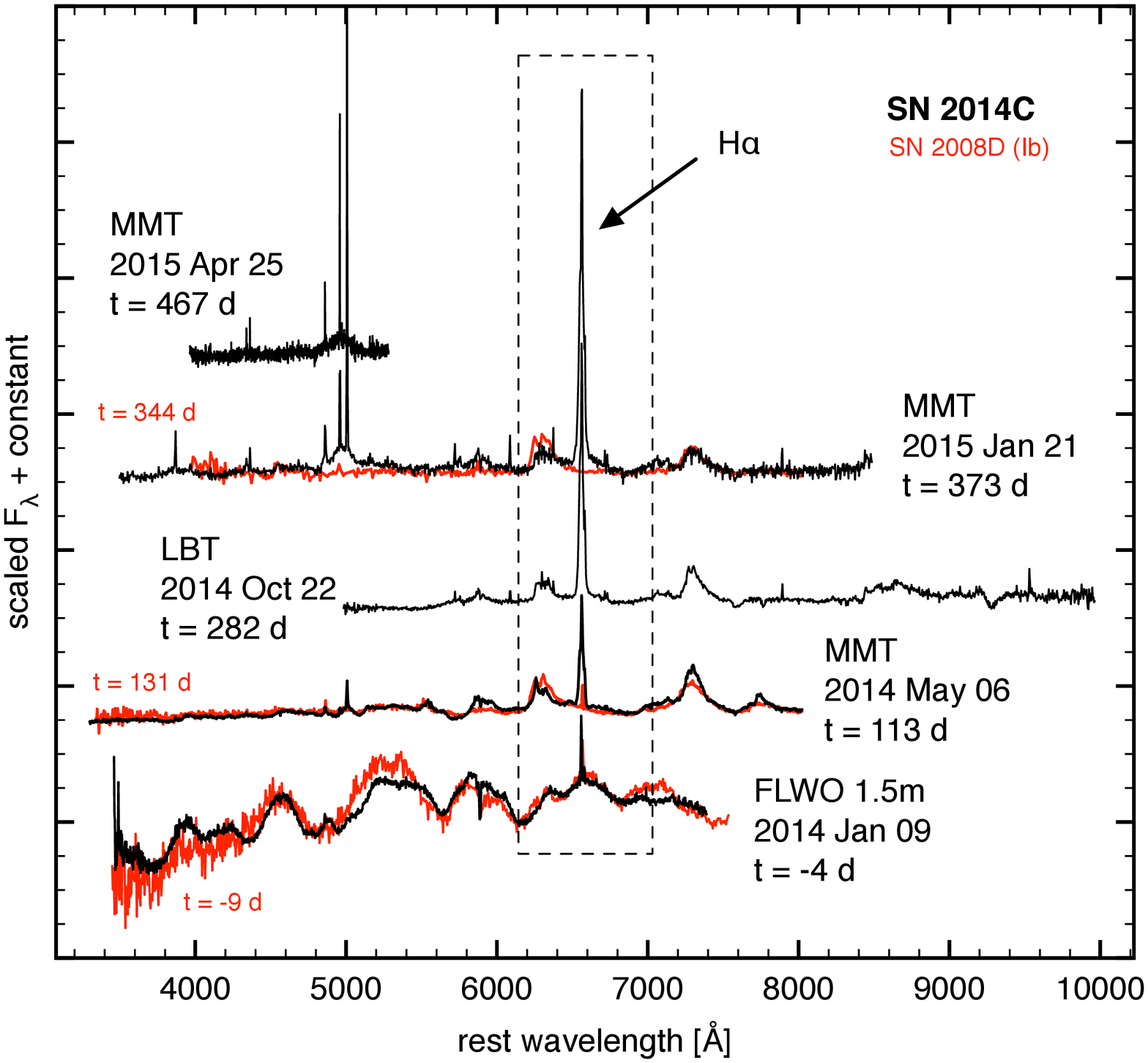}

\caption{Five epochs of optical spectra of SN\,2014C. Telescopes used,
  dates of observations, and phases with respect to V-band maximum on
  2014 January 13 are provided. Three epochs of spectra of the H-poor
  type Ib SN\,2008D, originally published in \citet{Modjaz09} and
  \citet{Tanaka09}, are shown for comparison. Epochs for SN\,2008D are
  with respect to V-band maximum on 2008 January 28
  \citep{Modjaz09}. At maximum light, SN\,2014C exhibits a spectrum
  very similar to that of SN\,2008D. One year later, however, the
  spectrum of SN\,2014C is largely dominated by H$\alpha$ with a
  profile having an extended FWHM width of $\approx 1400$ \kms\ that
  is normally only observed in H-rich type IIn supernovae. Refer to
  Figure~\ref{fig:lineID} for an enlargement of the day 373 spectrum.}

\label{fig:spectra}
\end{figure*}

\subsection{Optical Spectroscopy}
\label{sec:spectra}

Low-resolution optical spectra of SN\,2014C were obtained from three
telescopes: The F.~L.\ Whipple Observatory (FLWO) 1.5\,m Tillinghast
telescope mounted with the FAST instrument \citep{Fabricant98}, the
6.5\,m MMT Telescope mounted with the Blue Channel instrument
\citep{Schmidt89}, and the $2\times8.4$\,m Large Binocular Telescope
mounted with the MODS instrument \citep{Pogge10}. Details of the
observations are provided in Table~\ref{tab:speclog}.

\begin{figure}[htp!]
\centering
\includegraphics[width=\linewidth]{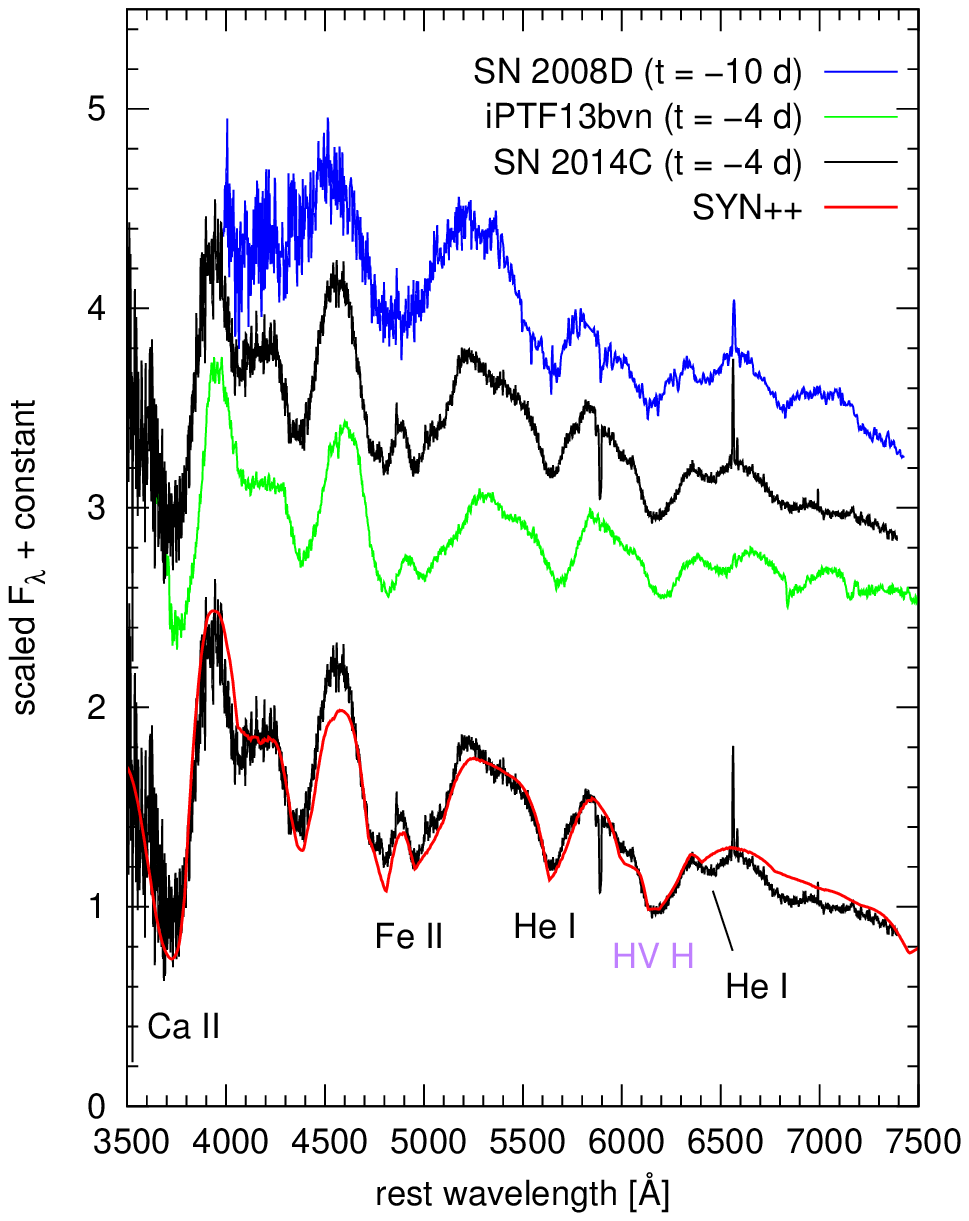}

\caption{The spectrum of SN 2014C compared to those of other type Ib
  supernovae near maximum light. The spectra of SN\,2014C and
  SN\,2008D have been corrected for extinction using $E(B-V)$ values
  of 0.75 mag (this paper) and 0.5 mag \citep{Soderberg08,Modjaz09},
  respectively, whereas the spectrum of iPTF13bvn has not been
  corrected \citep{Cao13,Srivastav14}. The data were originally
  published in \citet{Modjaz09} and \citet{Srivastav14} and were
  digitally retrieved from WISEREP \citep{Yaron12}. Also shown is a
  synthetic spectrum of SN\,2014C created with \texttt{SYN++}. Some
  absorption features dominated by single ions with projected Doppler
  expansion velocities of $(1.3 \pm 0.1) \times 10^{4}$ \kms\ are
  identified, as well as high velocity (HV) hydrogen that spans
  $(1.3-2.1) \times 10^{4}$ \kms.}

\label{fig:syn++}
\end{figure}

Our five epochs of optical spectra of SN\,2014C are plotted in
Figure~\ref{fig:spectra}. The explosion date is not tightly
constrained by the light curve, so we use the peak in the V-band on
2014 January 13 as the reference from which phase in days is measured
(Margutti et al.\ 2015).

Standard procedures to bias-correct, flat-field, and flux calibrate
the data were followed using the IRAF/PYRAF software\footnote{PYRAF is
  a product of the Space Telescope Science Institute, which is
  operated by AURA for NASA.} and our own IDL routines. A recession
velocity of 990 \kms, determined from many narrow emission lines
including [\ion{O}{3}] $\lambda\lambda$4959, 5007, H$\alpha$, and
[\ion{S}{2}] $\lambda\lambda$6716, 6731, was removed from all
spectra. Line identifications and estimates of expansion velocities of
the photospheric spectra were made with the supernova spectrum
synthesis code \texttt{SYN++} \citep{Thomas11}. Narrow emission line
identifications in late-time spectra were made using lists provided in
\citet{Fesen96}.

The foreground extinction due to the Milky Way is $E(B-V)_{mw} = 0.08$
mag \citep{Schlafly11}. The host internal extinction was estimated by
measuring the equivalent width (EW) of \ion{Na}{1}\,D absorption in
our optical spectra and following the prescriptions of
\citet{Turatto03}. The EW(NaID) from the day $-4$ spectrum is $4.25
\pm 0.07$ \AA. Using the lower branch of the \citet{Turatto03}
relation (see their Fig.\ 3), this measurement of EW(\ion{Na}{1})
implies $E(B-V)_{\rm host} = 0.67 \pm 0.01$ mag. We adopted a total
extinction of $E(B-V)_{\rm total}=0.75 \pm 0.08$ mag, which combines
the Galactic extinction with the inferred host extinction. We
confirmed that this extinction estimate provides an appropriate
$(B-V)$ color correction to the color indices of SN\,2014C to match
those of other type Ib supernovae (Margutti et al.\ 2015). All
extinction corrections made in this paper use $E(B-V)_{\rm total}$, in
combination with the standard reddening law of \citet{Cardelli89}
assuming $R_{\rm V} = 3.1$.

\subsection{Spectroscopic Metamorphosis}
\label{sec:morph}

Our optical spectrum obtained 2014 January 09 (day $-4$) shows
features clearly associated with \ion{Fe}{2}, \ion{He}{1},
\ion{Ca}{2}, exhibiting velocities of $(1.3 \pm 0.1) \times
10^4$~\kms\ (Figure~\ref{fig:syn++}). These are features regularly
seen in type Ib supernovae \citep{Filippenko97}, including well-known
examples SN\,2008D \citep{Modjaz09} and iPTF13bvn
\citep{Srivastav14}. An absorption centered around 6150~\AA\ is
typically associated with \ion{Si}{2}, however we find that the
spectrum is best fit with a combination of high velocity (HV)
H$\alpha$ features spanning $(1.3-2.1) \times 10^4$~\kms. This choice
of line identification is discussed in greater detail in
Section~\ref{sec:hydrogen}.

Follow-up spectroscopic observations did not resume until day 113 when
the supernova returned from behind the Sun. Interestingly, the day 113
spectrum of SN\,2014C exhibits a mix of standard and non-standard
emissions. The broad [\ion{O}{1}] $\lambda\lambda$ 6300, 6364 and
[\ion{Ca}{2}] $\lambda\lambda$7291, 7324 emission observed in the
spectrum are a normal feature of type Ib supernovae several months
after explosion and are associated with inner metal-rich ejecta that
are radioactively heated by $^{56}$Co. Most unusual, however, was
conspicuous emission centered around the H$\alpha$ line with an
overall FWHM of 1400 \kms\ that had emerged. This emission continued
to grow in strength relative to other emission lines over the next
several months (Figure~\ref{fig:spectra}).

{\it An extraordinary event must have occurred while SN\,2014C was
  hidden behind the Sun.} An intermediate-width H$\alpha$ feature is
normally seen only in type IIn supernovae, where it is associated with
radiative shocks in dense clouds \citep{Chugai94}.  The interaction
decelerates the blast wave and a dense shell traveling approximately
at the shock velocity is formed. Accordingly, we interpret the
conspicuous change in H$\alpha$ emission from SN\,2014C to be the
result of the supernova having encountered dense H-rich circumstellar
material (CSM) between Februrary and May 2014. Consistent with this
scenario (see, e.g., \citealt{Chevalier06}), strong radio and X-ray
emission accompanied the sudden increase in H$\alpha$ emission in
SN\,2014C as the shock continued to strongly interact with the density
spike in CSM (Margutti et al. 2015; Kamble et al.\ 2015).

\begin{figure}[htp!]
\centering
\includegraphics[width=0.8\linewidth]{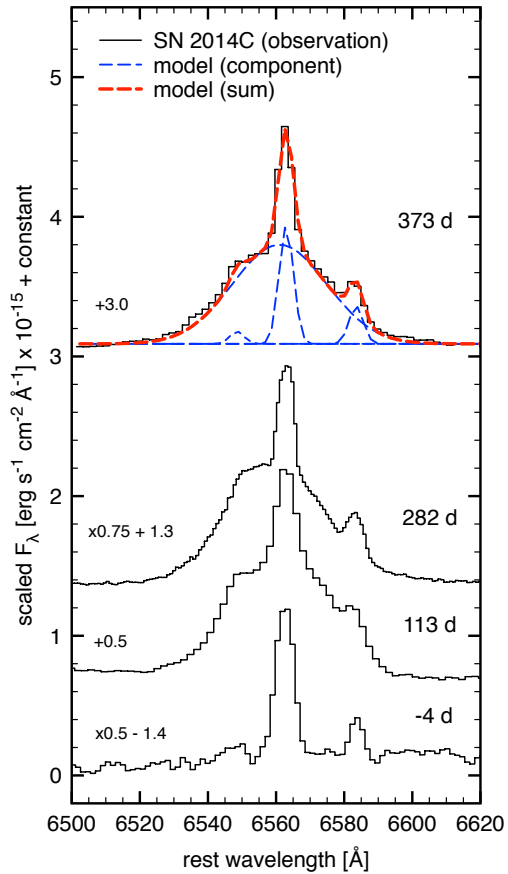}

\caption{Evolution of the emission line profile around H$\alpha$. On
  day $-4$ only unresolved (FWHM $\approx 250$ \kms)
  H$\alpha$ and [\ion{N}{2}] $\lambda\lambda$6548, 6583 lines are
  observed. Subsequently, beginning on day 113, an intermediate
  component (FWHM $\approx 1200$ \kms) is observed. A model fitting
  the day 373 profile is shown. Scaling parameters
  are given to the left of each spectrum.}

\label{fig:profile}
\end{figure}

We attempted to decompose the complex line profile centered on
H$\alpha$ in the spectra into multiple Gaussian features. After
removing the linear continuum and running a least-squares fitting
routine, we found that the profile could be reasonably reproduced with
narrow components of H$\alpha$ and [\ion{N}{2}] 6548, 6583 lines
having instrumentally unresolved FWHM velocities of $\approx 250$
\kms\ and an intermediate-width component with a FWHM width of $1200
\pm 100$ \kms\ (Figure~\ref{fig:profile}). The H$\alpha$ profile
extends to 6520 \AA\ and 6610 \AA, which sets limits on the velocity
of the emitting shocked CSM to $-2000$ and $+2200$ \kms. The narrow
components are presumably associated with wind material that is being
photoionized by X-rays of the forward shock, and the intermediate
component is associated with the shock and/or ejecta running into
CSM. The day $-4$ spectrum, which has only narrow components, is most
likely a combination of emission local to the supernova and from the
entire host massive star cluster (see
Section~\ref{sec:pre-explosion}). We interpret emission at later
epochs to be be dominated by emission from supernova-CSM interaction.

\begin{figure*}
\centering
\includegraphics[width=\linewidth]{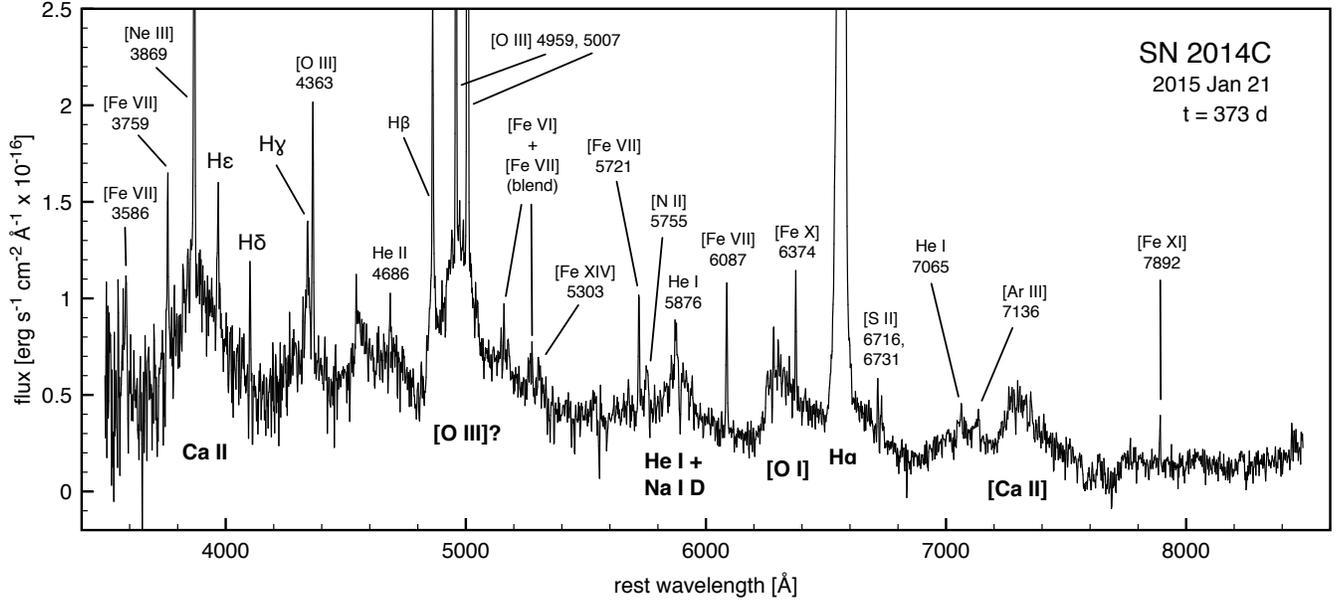}
\caption{Dereddened MMT spectrum of SN 2014C. The emission arises from
  various regions of shocked and photoionized CSM and stellar
  ejecta. Broad components are identified in bold below the spectrum,
  and narrow components are identified above the spectrum.}

\label{fig:lineID}
\end{figure*}

Beginning with the day 282 spectrum and continuing with the day 373
spectrum, the emissions are increasingly complex and originate from
several distinct regions. Figure~\ref{fig:lineID} shows an enlargement
of the day 373 spectrum corrected for extinction and a complete list
of identified emission features. Several narrow, unresolved emission
lines are observed including [\ion{O}{3}] $\lambda$4363 and
$\lambda\lambda$4959, 5007, [\ion{Ne}{3}] $\lambda$3869, \ion{He}{2}
$\lambda$4686, and [\ion{N}{2}] $\lambda$5755. Also seen are several
narrow, unresolved coronal lines including [\ion{Fe}{6}],
[\ion{Fe}{7}], [\ion{Fe}{10}], [\ion{Fe}{11}], and [\ion{Fe}{14}]. We
attribute this emission to ionization of the pre-shock circumstellar
gas by X-rays emitted by the shocked gas. The strongest constraint on
the wind velocity comes from the day 474 spectrum that has the highest
resolution of all our data. We measure a FWHM width of 1.5 \AA\ (which
is unresolved) in the [\ion{O}{3}] $\lambda$4363 emission line. This
sets an upper limit of $< 100$ \kms\ for the unshocked wind velocity.

Broad emission centered around the [\ion{O}{3}] $\lambda\lambda$4959,
5007 lines is seen in our spectra beginning on day 282 and continues
to be visible through our last spectrum obtained on day 474. The width
of the emission is difficult to measure since it blends with H$\beta$
blueward of 4959~\AA\ and another source of emission redward of
5070~\AA. We estimate that the velocity width must be $\ga 3500$ \kms,
meaning that the emission originates from a region different than the
shocked CSM. Presumably it is emission from oxygen-rich stellar ejecta
being excited by the reverse shock. Broad [\ion{O}{3}] emission is
normally only seen in supernovae many years to decades after core
collapse \citep{Milisavljevic12}. But in the case of SN\,2014C, the
supernova-CSM interaction may have accelerated its dynamical
evolution.

By comparing the overall emission profile around 7300 \AA\ to that
observed around [\ion{O}{1}] $\lambda\lambda$6300, 6364, we find that
the velocity distribution is best matched when measured with respect
to the [\ion{Ca}{2}] $\lambda\lambda$7291, 7324 lines ($\lambda_C =
$7307 \AA) as opposed to the [\ion{O}{2}] $\lambda\lambda$7319, 7330
lines ($\lambda_C = $7325 \AA). Detecting emission from [\ion{O}{1}]
and [\ion{Ca}{2}], but not from [\ion{O}{2}], implies that the ejecta
associated with this emission have densities that are $\ga 10^{6}$
cm$^{-3}$ \citep{Fesen99}.

\begin{figure*}
\centering
\includegraphics[width=0.8\linewidth]{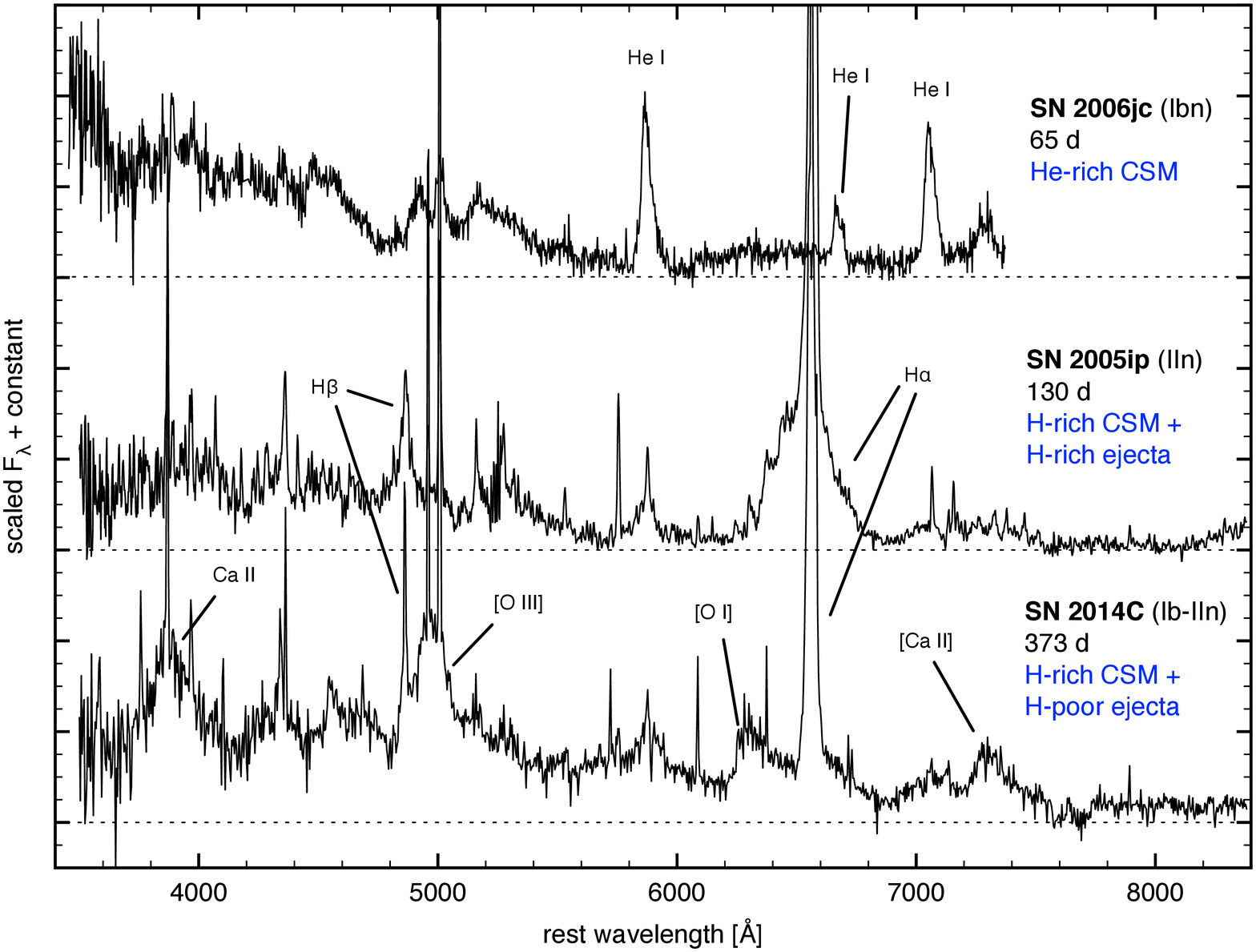}
\caption{Dereddened MMT spectrum of SN 2014C compared to the spectra
  of the type IIn SN\,2005ip \citep{Stritzinger12} and the type Ibn
  SN\,2006jc \citep{Modjaz14} retrieved from WISEREP. Prominent
  emission features associated with excited CSM and supernova ejecta
  are highlighted. Type IIn supernovae exhibit complex H Balmer
  profiles that are a blend of broad ($>5000$ \kms), intermediate
  ($\sim 1000$ \kms) and narrow ($<200$ \kms) velocity components
  associated with H-rich ejecta, supernova-CSM interaction, and
  photoionized CSM, respectively. Type Ibn supernovae exhibit
  prominent intermediate width \ion{He}{1} emission from supernova-CSM
  interaction and no emission from interior ejecta. SN\,2014C exhibits
  intermediate width H$\alpha$ and emission from interior O-rich
  material excited by radioactive decay of $^{56}$Co. Many narrow
  lines associated with photoionized wind material are shared between
  all three supernovae. Dashed horizontal lines provide a baseline to
  compare increases in continuum strength toward shorter wavelengths.}

\label{fig:IInIbn}
\end{figure*}

The dereddened day 373 optical spectrum exhibits a blue
pseudo-continuum, which is sometimes seen in type IIn and Ibn
supernovae and has been attributed to fluorescence from a number of
blended emission lines (\citealt{Fransson02,Foley07,Smith09};
Figure~\ref{fig:IInIbn}). One of the strongest expected
fluorescence-pumped \ion{Fe}{2} lines is at 8451 \AA\
\citep{Fransson02}. Near this wavelength region we detect a minor
emission peak centered at 8446~\AA\ in our day 282 spectrum that is
blended with the \ion{Ca}{2} near-infrared triplet. This feature is
most likely the \ion{O}{1} recombination line at 8446 \AA\ and not
\ion{Fe}{2}.  We also do not observe Fe II lines at 9071, 9128, or
9177 \AA, which would be expected to accompany the Fe II $\lambda$8451
line \citep{Fransson02}.

\subsection{Properties of the unshocked CSM}

Properties of the surrounding unshocked CSM shed from the progenitor
star of SN\,2014C can be estimated using the relative line strengths
of narrow features observed in our optical spectra. The forbidden
oxygen lines provide a lower limit to the density. We do not detect
narrow [\ion{O}{2}] $\lambda$3727, but we do detect [\ion{O}{3}]
$\lambda\lambda$4959, 5007, indicating that electron densities are
well above $10^4$ cm$^{-3}$ in this emitting region
\citep{Osterbrock06}. An upper limit to the density can be estimated
from the relative line strengths of [\ion{Fe}{7}]. Comparing these
line strengths with the CHIANTI database \citep{Landi13}, we find that
the density is less than $10^7$ cm$^{-3}$.  An estimate of the
temperature can be derived using the [\ion{O}{3}] line diagnostic $R =
\lambda$(4959+5007)/$\lambda$4363 for densities between $10^{5} -
10^{6}\;\rm cm^{-3}$ using the \texttt{TEMDEN} task in IRAF
\citep{Shaw94}. We measure $R = 9.8 \pm 0.2$, which is associated with
temperatures between $(2-8) \times 10^4$\,K.

The ratio of [\ion{N}{2}] $\lambda$6583 / H$\alpha \approx 0.3$ in the
day 373 spectrum is higher than the ratio observed in the day $-4$
spectrum (i.e., before interaction commenced) where it is $\approx
0.15$. An increase in temperature, which would be anticipated from the
hard ionizing spectrum and high density that suppresses some of the
forbidden line cooling, may explain the high ratio. The increase could
also be indicative of nitrogen-enriched CSM from CNO processing in the
progenitor star.

Knowledge of the X-ray luminosity $L_x$ around day 373 can further
constrain properties of the unshocked CSM. The ionization parameter is defined as
\begin{equation}
\xi = L_x/(n r^2),
\end{equation}
where $n$ is the electron density number, and $r$ is the radius of the
emitting region. $L_x$ at this time is $\sim 5 \times 10^{40}$ erg
s$^{-1}$ and consistent with a temperature of 18 keV (Margutti et al.\
2015).  \citet{Kallman82} find that the \ion{Fe}{7} ion fraction peaks
at about 30\% for $\xi \sim 10$ in models with photoionizing 10 keV
bremsstrahlung spectra that are expected to be at least
approximately applicable to SN\,2014C. Thus,
\begin{equation}
n r^2 = L_x/\xi \sim 5 \times 10^{39}\;\rm cm^{-1}
\label{eq:nr2}
\end{equation}
Although there must be a range of ionization parameters present, the
ratio of [\ion{Fe}{14}] $\lambda$5303/[\ion{Fe}{10}] $\lambda$6374 is
only $0.18 \pm 0.03$, which indicates that $\xi$ only goes up to
$\approx 25$.

The emission measure $EM$ is defined as
\begin{equation}
\begin{split}
  EM & = (4 \pi/3) n^2 r^3 f \\
  & = \frac{L_{\rm [Fe~VII]} f}{  j_{\epsilon} A_{(Fe)} \eta}
\label{eq:EM1}
\end{split}
\end{equation}
where $j_{\epsilon}$ is the emissivity, $A_{(Fe)}$ is the Fe abundance
($2.75\times 10^{-5}$), $\eta$ is the ion fraction, and $f$ is the
filling factor. We measure the line flux of [\ion{Fe}{7}]
$\lambda$6087 to be $5.2 \times 10^{-15}$ erg cm$^{-2}$ s$^{-1}$,
which translates to a luminosity of $1.4 \times 10^{38}$ erg
s$^{-1}$. The emissivity for the [\ion{Fe}{7}] $\lambda$6087 line from
CHIANTI is $1.3 \times 10^{-20}$ erg\,s$^{-1}$\,cm$^{-3}$\,sr$^{-1}$.

Solving equation (\ref{eq:EM1}) for $r$ in terms of equation (\ref{eq:nr2}),
\begin{equation}
r = \frac{(n r^2)^2}{EM(4 \pi/3)f},  
\end{equation}
yields a radius of $ 4.9\times 10^{15}\;f^{-1}\;\rm cm$.
Recasting equation (\ref{eq:nr2}) in terms of $n$, and substituting this
value of $r$ we find that
\begin{equation}
n = L_x/(\xi r^2) \sim 2 \times 10^8\;f^2\;\rm cm^{-3}.
\label{eq:n}
\end{equation}
With the constraint that $n \la 10^{6}\;\rm cm^{-3}$ from the
forbidden oxygen and iron line diagnostics, equation (\ref{eq:n})
implies that $f \la 0.1$, which is reasonable since the gas that
contributes to the highest-ionization Fe lines may fill only some of
the volume. The temperatures, densities, and filling factor
we derive suggest that the emission is not uniformly distributed
and originates from clumped material. A full treatment of emission
line modeling is left for future work.

\section{Discussion}
\label{sec:Discussion}

SN\,2014C represents the first time an ordinary type Ib supernova has
been seen to slowly evolve into a strongly interacting type IIn. This
Ib-IIn metamorphosis is consistent with a delayed interaction between
an H-poor star's supernova explosion and a local H-rich shell
presumably formed from material stripped from the progenitor star. The
strong supernova-CSM interaction must have started between February
and May of 2014 while the supernova was behind the Sun. Type Ib
supernovae have typical rise times of $\la 20$ days
\citep{Drout11,Bianco14}, thus the strong interaction occurred
somewhere between $\approx$ 20 and 130 days after explosion. Assuming
a minimum forward shock velocity of $0.1\,c$, we can loosely constrain
the radius of the shell to be $\ga 3 \times 10^{16}$ cm from the
explosion center.

In Figure~\ref{fig:IInIbn}, SN\,2014C is compared to two
representative objects: the type IIn SN\,2005ip \citep{Stritzinger12}
and the type Ibn SN\,2006jc \citep{Modjaz14}. Spectroscopic
differences and similarities between these supernovae reflect the
distance of the CSM shell with respect to the explosion and the
shell's chemical composition. In type IIn and Ibn supernovae, massive
and dense CSM is located nearby and interaction commences shortly
($\la 1$ day) after explosion. Consequently, an opaque CSM hides the
interior and O-rich material is often never observed. This contrasts
with SN\,2014C, where the longer delay between explosion and
interaction is associated with a larger, less dense and less opaque
shell. This latter situation is advantageous as it permits
observations of interior processes that are normally hidden.

\subsection{Previous examples of delayed interaction} 

The closest analog to SN\,2014C in overall properties is
SN\,2001em. That supernova was spectroscopically classified as a
hydrogen deficient type Ib/c \citep{Filippenko01}\footnote{The text of
  \citet{Filippenko01} reads ``type Ib or Ic (most likely Ic), perhaps
  a month after maximum brightness.''  We visually inspected these
  data published in \citet{vanDyk10} and agree with their
  classification.}, two years before it was re-detected as a highly
luminous radio source \citep{Stockdale04}, and showed
intermediate-width H$\alpha$ emission with FWHM $\sim 1800$ \kms\ in
its optical spectra \citep{Soderberg04}. Unlike SN\,2014C, the
intermediate stages between H-deficient to H-dominated emission in
SN\,2001em was completely missed.

Although it was originally thought that SN\,2001em may harbor a
bipolar relativistic off-axis jet that had decelerated to mildly
relativistic velocities \citep{Granot04}, later observations with very
long baseline interferometry ruled this model out by demonstrating
that the expansion velocity of the supernova was below 6000 \kms\ and
that the supernova was therefore not driven by a relativistic jet
\citep{Bietenholz05,Bietenholz07,Schinzel09}. \citet{Chugai06} found
that the multi-wavelength properties of SN 2001em could be modeled in
a scenario in which non-relativistic ejecta from the supernova
explosion collided with a dense and massive ($\sim 3\;M_{\odot}$)
circumstellar shell at a distance of $\sim 7 \times 10^{16}$ cm from
the star. The circumstellar shell was presumably formed by a vigorous
mass loss episode with a mass loss rate of $\sim 2 \times 10^{-3}$
M$_{\odot}$ yr$^{-1}$ approximately $1000$ to $2000$ yr prior to the
supernova explosion. The hydrogen envelope was completely lost and
subsequently swept up by the fast wind of the pre-supernova star and
accelerated to a velocity of 30-50 \kms.  Such a high-rate mass-loss
event could be explained by binary interaction, but could also be
explained by a powerful eruption from an LBV star. A similar scenario
is possible for SN\,2014C, but in this case, less of the progenitor's
H-rich envelope had been stripped at the time of explosion (see
additional discussion in section \ref{sec:shell}).

Late-time interaction with dense CSM shells has been observed in a
handful of supernovae with H-rich ejecta. For example, the progenitor
of SN\,1996cr evacuated a large cavity just prior to exploding, and
the forward blast wave likely spent 1-2 yr in relatively uninhibited
expansion before eventually striking a dense shell of CSM
\citep{Bauer08}. SN\,1996cr may be a ``wild cousin'' of SN\,1987A,
which has slowly interacted with a dense ring at a radius of $6 \times
10^{17}$ cm (approximately an order of magnitude larger than the shell
radius inferred for SN\,2014C) beginning as early as 1995
\citep{Sonneborn98,Lawrence00}.

\subsection{Partially stripped progenitor star}
\label{sec:hydrogen}

Our modeling of the near-maximum light optical spectrum of SN\,2014C
associates absorption around 6150~\AA\ with high velocity hydrogen
(Figure~\ref{fig:spectra}). Attempts to model the 6150~\AA\ absorption
with \ion{Si}{2} were unsuccessful because the velocity as inferred
from the minimum of the feature ($v \approx 8000$ \kms) was
inconsistent with those inferred from all other ions ($v \approx
13000$ \kms, cf.\ Section \ref{sec:morph}). Alternative
identifications for this absorption feature that have been
investigated elsewhere for other SN Ibc, including \ion{C}{2} and
\ion{Ne}{1} \citep{Harkness87,Deng00}, were ruled out as strong
contributors in SN\,2014C.  Our model is consistent with recent work
by \citet{Parrent15}, who find that \ion{Si}{2} is unlikely to be the
dominant contributor to the absorption feature near 6150~\AA\ seen in
many SN\,Ibc.

The detection of hydrogen absorption around the time of maximum light
implies that the progenitor star of SN\,2014C was only partially
stripped of hydrogen at the time of explosion. This is an important
observation because it informs about the evolutionary status of the
star at the time of core collapse. It is also relevant for debates on
the extent of H and He in SN\,Ibc
\citep{Matheson01,Branch02,Branch06,Hachinger12,
  Milisavljevic13,Modjaz14,Milisavljevic15,Parrent15}.  The range of
projected Doppler velocities used in fitting the optical spectrum with
\ion{H}{1} ($1.3-2.1 \times 10^4$~\kms) is consistent with a detached
envelope consisting of an extended layer of hydrogen beyond the
otherwise fairly sharp photosphere. It is beyond the scope of this
paper to accurately estimate the mass of hydrogen associated with the
detached envelope, but the models of \citet{Hachinger12} provide
useful upper limits. Hachinger et al.\ conclude that $\approx 0.03\;
M_{\odot}$ of hydrogen is sufficient for H$\alpha$ absorption to
dominate over absorption due to \ion{Si}{2} $\lambda$6355, as it does
in SN\,2014C.

Mass loss in massive stars is correlated with metallicity
\citep{Vink01,Vink05}. The oxygen abundance of the explosion site of
SN\,2014C measured from our day $-4$ spectrum using the $N2$ scale of
\citet{Pettini04} is $\log({\rm O/H})+12 = 8.6 \pm 0.1$, which is near
the solar value ($\log({\rm O/H})_{\odot} + 12 = 8.69$;
\citealt{Asplund05}). Our measurement is consistent with an
independent measurement of the oxygen abundance of the center of
NGC\,7331 ($\log({\rm O/H}) + 12 = 8.75 \pm 0.18$; \citealt{Gusev12}).
The explosion site metallicity of SN\,2014C is above the median
metallicity of type Ib supernovae with secure classifications, which
is $\log({\rm O/H})_{\odot} + 12 = 8.43 \pm 0.14$ \citep{Sanders12},
and in line with the mean metallicity of type IIn, which is $\log({\rm
  O/H})_{\odot} + 12 = 8.63 \pm 0.03$ \citep{Taddia15} (both measured
with the same $N2$ scale).  However, the explosion site metallicity is
consistent with the metallicity distribution observed for each class,
which are largely overlapping. SN 2014C was discovered in a targeted
survey, which is known to favor the discovery of supernovae in
brighter and more metal-rich galaxies \citep{Sanders12}.

\subsection{Pre-explosion detection of progenitor system}

Valuable information about the progenitor system comes from the Subaru
and HST pre-explosion imaging (Figures~\ref{fig:images} and
\ref{fig:enlarge}). Correcting for extinction using $E(B-V)_{\rm
  total}$, the absolute magnitude of the coincident source is $M_V =
-12.0 \pm 0.17$ mag. The luminosity rivals that of the most luminous
stars known, suggesting that the source is not a single star and is
instead a massive star cluster. Supporting this conclusion are the
facts that 1) the source is extended and composite in the HST F658N
image with an effective radius of $\approx 17$ pc (cf.\ Section
\ref{sec:pre-explosion}) that is consistent with observed sizes of
massive star clusters (see, e.g., \citealt{Bastian13}), and 2)
SN\,2014C has a noticeable offset from the center of the source
(Fig.~\ref{fig:images}).

We compared the photometry and estimate of H$\alpha$ luminosity of the
pre-explosion source to the Binary Population and Spectral Synthesis
(BPASS) stellar population models (\citealt{Eldridge09}; Eldridge et
al., in preparation; http://bpass.auckland.ac.nz). Assuming the total
extinction $E(B-V)_{\rm total}$, we find that the best fitting age of
the stellar population is between 30 to 300 Myr. The detection of
H$\alpha$ emission from the pre-explosion source favors younger
populations closer to 30 Myr, and binary star models match estimates
derived from the H$\alpha$ luminosity more closely than those derived
from single star models.  The turn off mass of stellar populations in
the favored age range is between 3.5 to $9.5\;M_{\odot}$, and the mass
of the cluster is estimated to be between $3 \times 10^5$ and $10^6
M_{\odot}$. These age constraints are the same for three
metallicities close to that of the environment of $Z=0.008$, $0.014$
and $0.020$, where solar metallicity is between $0.014$ and $0.020$.

A source of uncertainty in our analysis of the host massive star
cluster is in our estimate of the foreground extinction. It is unknown
whether the extinction estimated from the day $-4$ optical spectrum of
SN\,2014C is local to the supernova, or along the line of sight to the
entire cluster. An additional complication is that if the extinction
is local, then it is possible that the extinction may have changed
with time. Thus, within the stated uncertainties, the uncorrected
absolute magnitudes represent lower limits to the luminosity of the
cluster.

\subsection{Origin of the CSM shell}
\label{sec:shell}

Massive stars are known to become stripped of their outer layers in a
variety of ways including line-driven winds and binary interactions
\citep{Podsiadlowski92,Woosley95,Wellstein01,Vink01,Puls08,Yoon10},
but the details behind the precise physical mechanisms that are
involved and the stages that this mass loss takes place are not well
constrained (see \citealt{Eldridge08,Langer12,SmithARAA14} and
references therein). Any plausible explanation of the origin of the
dense CSM shell with which SN\,2014C interacted must be able to
account for two key properties of the system. One property is the
distance of the shell from the progenitor star of SN\,2014C, which is
much more distant than the shells in the majority of type IIn and Ibn
supernovae. Another property is that the progenitor star exploded as a
type Ib supernova, meaning that it had been largely stripped of its
hydrogen envelope at the time of core collapse.

Below, we discuss three plausible scenarios for the physical origin of
the massive shell that surrounded SN\,2014C at the time of explosion.

\subsubsection{An unusually short W-R phase}

W-R stars have significant stellar winds, with mass loss rates of a
few times $10^{-5} M_{\odot}$ yr$^{-1}$ and terminal wind velocities
of $1000-3000$ \kms \citep{Crowther07}. These winds can run into and
overtake the much slower winds ($< 100$ \kms) of a previous red
supergiant (RSG) phase and create massive CSM shells, or ``bubbles,''
(see, e.g., \citealt{Garcia96}). Approximately half of the Galactic
population of W-R stars are associated with a CSM shell, some of which
have anomalous abundances associated with ejected CNO-processed
stellar material \citep{Miller93}. However, the majority of these CSM
shells have diameters of many parsecs, which is much larger than the
radius of the shell inferred for SN 2014C ($\sim 0.01$ pc). Thus,
SN\,2014C did not interact with an environment like those typically
observed around W-R stars.

However, it is possible that the progenitor star of SN\,2014C evolved
through a brief W-R phase lasting much less than $10^4$ yr. Although
traditionally it has been believed that massive stars should spend
$0.5-1$ Myr in the core-He-burning W-R phase before finally exploding
as normal SN Ibc \citep{Heger03}, advances in stellar evolution
modeling show many pathways that may potentially lead to short-lived
W-R stars. Examples include low mass binary stars \citep{Eldridge08},
and RSGs with strong pulsation-driven superwinds
\citep{Yoon10-RSG}. Analyses of circumstellar environments around
supernova remnants also support the notion that progenitors of
stripped-envelope supernovae may pass through brief W-R phases
\citep{Schure08,Hwang09}.

However, a short W-R phase associated with a typical mass loss rate of
a few $\times 10^{-5}\;M_{\odot}\,\rm yr^{-1}$ in itself cannot
explain the massive shell SN\,2014C interacted with. Though a short
W-R phase will sweep up a dense shell close to the star, the shell
will not have much mass because the RSG wind was expanding freely at
this small radius. A possible solution is if the short W-R phase was
accompanied with an enhanced mass loss rate (e.g., $\sim
10^{-3}\;M_{\odot}\;\rm yr^{-1}$), which is a scenario proposed by
\citet{Chugai06} for SN\,2001em.

\subsubsection{Shell ejection in a single eruption}

Alternatively, in light of the established connection between type IIn
supernovae and LBV stars, it is plausible that the shell surrounding
the SN\,2014C progenitor system was ejected in a single eruptive event
rather than a brief period of enhanced W-R stellar wind. Shell
ejections have typical velocities of $\ga 100$ \kms, which would mean
that the major mass loss event would have taken place in the last $\la
100$ years prior to core collapse.

Providing the pre-explosion source was a massive star cluster, the
most probable turn-off mass range estimated from the BPASS models
($3.5 - 9.5\; M_{\odot}$) is too low to be compatible with typical LBV
progenitors ($M_{ZAMS} \ga 20 M_{\odot}$). However, because
uncertainties in our fitting allow room for ages as low as 10~Myr, an
LBV-like progenitor is not ruled out completely. It is worth noting
that the type IIn-LBV connection is thus far only robust for stars
with H-rich envelopes and to date there is no direct detection of the
progenitor star of an interacting H-poor supernova. Indirect arguments
that the progenitors of type Ibn supernovae exploded in the transition
from LBV to W-R phases \citep{Foley07,Pastorello07,Smith12},
potentially due to instabilities initiated in the final stages of He
core burning \citep{Pastorello15}, have been advanced.

Ejection of an H-rich common envelope in a binary system is also a
plausible scenario. Stellar evolution calculations and population
models that incorporate stellar duplicity show that stars less than 20
M$_{\odot}$ in interacting binaries can be stripped of their hydrogen
envelopes via mass transfer to a companion and/or the loss of the
common envelope, and end their lives as low mass helium stars that
explode as SN Ibc \citep{Yoon10,Eldridge13}. This type of progenitor
system was proposed for the type Ib iPTF13bvn \citep{Eldridge15},
which has an optical spectrum at maximum light very similar to that of
SN\,2014C (see Figure~\ref{fig:syn++}).

\subsubsection{CSM confinement}

The location of SN\,2014C's progenitor star within a compact massive
star cluster makes it plausible that hot gas from stellar winds and
prior supernovae could have provided a large external pressure by
which to confine the RSG wind into a shell near the star.  Simulations
have shown that it is difficult to confine RSG wind to $<0.3$ pc with
external thermal or ram pressure \citep{vanMarle06,Eldridge06}, thus
this scenario is not plausible unless the wind had considerable
asymmetry originating from stellar rotation and/or duplicity
\citep{Eldridge07}.

A viable alternative scenario is through the process of
photoionization-confinement. \citet{Mackey14} demonstrate how an
external radiation field generated by neighboring stars can form a
standing shock in the neutral part of an outflowing wind and create an
almost static, photoionization-confined shell that traps up to 35 per
cent of all mass lost during the RSG phase close to the star until it
explodes. Their model was specific to Betelgeuse and its $\sim
0.1\;M_{\odot}$ circumstellar shell, but the model is applicable to
other stars that might have much more massive shells.

Complicating this interpretation is the fact that the progenitor star
of SN\,2014C was H-stripped at the time of explosion and thus no
longer in the RSG phase. If the star evolved through a W-R phase with
a typical timescale of $\sim 10^5$ yr, the momentum of the winds
traveling $\ga 1000$ \kms\ make it unlikely that a shell $\la
1\;M_{\odot}$ in mass with radius $\sim 3 \times 10^{16}$\,cm would
have survived. The photoionization-confined shell scenario remains
plausible if the shell was initially much more massive at the end of
the RSG phase (a few $\times 1\;M_{\odot}$) and/or if the W-R phase
was short.

\section{Conclusion}
\label{sec:Conclusions}

We have presented spectroscopic observations of SN\,2014C that follow
its evolution from an ordinary type Ib supernova to an interacting
type IIn. Our observations are consistent with the supernova having
exploded in a cavity before encountering a dense, nearby ($\ga 3
\times 10^{16}$ cm) H-rich shell formed from mass lost from the
progenitor star. We considered three possible origins to the shell: 1)
a W-R fast wind phase that overtook a slower RSG wind, 2) an eruptive
ejection, or 3) various forms of CSM confinement. We find that all
explanations require that the progenitor star experienced a brief $\la
1000$\,yr W-R phase. The brief W-R phase may have been associated with
larger-than-normal mass loss rates ($> 10^{-4}\;M_{\odot}\;\rm
yr^{-1}$). Alternatively, the prior RSG wind may have been distributed
asymmetrically and/or confined via a photoionization layer. Our
observations disfavor a sudden eruption in an LBV-like event, but do
not rule out such a scenario. Ejection of an H-rich common envelope in
a binary system is also possible.

We also presented archival Subaru and HST pre-explosion images
covering the field of SN\,2014C that show a luminous coincident
source. The SED and source size as measured by HST are both consistent
with a compact massive star cluster. We estimated the age of the
cluster to be $30 - 300$ Myr, and favor models incorporating
interacting binary systems with ages closer to $30$ Myr in light of
relatively strong H$\alpha$ emission. 

Because extensive archival data of the host galaxy exist, it is
possible that a historical light curve could be constructed. Such a
light curve could be used to look for changes in apparent brightness
in the system that may be associated with precursor activity from an
unstable massive star (see, e.g., \citealt{Pastorello13} and
\citealt{Ofek14}). Notably, SN\,2013bu was discovered in NGC\,7331 on
2013 April 21 \citep{Itagaki13}, which was only eight months before
the discovery of SN\,2014C. Thus, multi-epoch observations of
SN\,2013bu that cover the region of SN\,2014C would be appropriate for
this type of archival analysis.

We searched for previous HST observations, but unfortunately
SN\,2014C fell slightly outside the footprint except for the F658N
images presented here. Future observations can pinpoint the location
within the host cluster and investigate the immediate stellar
environment when the supernova has faded. Our results strongly
motivate observations at ultraviolet wavelengths from which line
diagnostics can be performed to investigate potentially enhanced
abundances consistent with CNO processed material (e.g.,
\citealt{Fransson02,Fransson14}). It is anticipated that additional
emission lines will develop as the CSM cools over time (see, e.g.,
SN\,2005ip; \citealt{Smith09}). Line diagnostics at these wavelengths
would reveal important information about the evolutionary transitions
a massive star may undergo in its final stages approaching core
collapse.

SN 2014C is a well-observed example of a class of core-collapse
supernovae that fill a gap between events that interact strongly with
nearby environments immediately after explosion (type IIn and Ibn) and
events that are never observed to interact at all (the majority of
SN\,Ibc). Previous surveys of late-time radio emission from SN\,Ibc
suggest that events like SN\,2014C are infrequent
\citep{Bietenholz14}. However, considering that these follow-up
observations are not densely sampled and cover many SN\,Ibc that are
$> 50$ Mpc, the frequency of other late interaction events where the
mass lost is less extreme and/or concentrated at greater distances is
unknown.

The shell of mass loss material surrounding the progenitor star of
SN\,2014C was close enough to be detected via subsequent interaction
with the blast wave, and yet fortuitously distant enough to permit a
clear view of the underlying supernova. Improved constraints on the
frequency with which SN\,Ibc interact with their stripped shells can
contribute important information about the final stages of mass loss
and stellar evolution. Given that the hydrodynamic instabilities that
lead to enhanced pre-supernova mass loss may be related to deviations
from spherical symmetry in the progenitor star structure
\citep{SmithArnett14}, understanding how the underlying physical
mechanisms may permit a range of pre-supernova mass loss rates and
time lags may also help with our understanding of the core collapse
process itself.

\acknowledgements

We thank an anonymous referee who provided many helpful comments and
suggestions that improved the quality and presentation of this
paper. Observations reported here were obtained at the MMT
Observatory, a joint facility of the Smithsonian Institution and the
University of Arizona, as well as the 6.5 m Magellan Telescopes
located at Las Campanas Observatory, Chile. This paper uses data taken
with the MODS spectrographs built with funding from NSF grant
AST-9987045 and the NSF Telescope System Instrumentation Program
(TSIP), with additional funds from the Ohio Board of Regents and the
Ohio State University Office of Research. Based in part on data
collected at Subaru Telescope and obtained from the SMOKA, which is
operated by the Astronomy Data Center, National Astronomical
Observatory of Japan. Some of the data presented in this paper were
obtained from the Mikulski Archive for Space Telescopes (MAST). STScI
is operated by the Association of Universities for Research in
Astronomy, Inc., under NASA contract NAS5-26555. Support for MAST for
non-HST data is provided by the NASA Office of Space Science via grant
NNX09AF08G and by other grants and contracts. This paper made use of
the Weizmann interactive supernova data repository (WISEREP) -
http://wiserep.weizmann.ac.il. J.~M.\ acknowledges support from the
Deutsche Forschungsgemeinschaft priority program 1573, Physics of the
Interstellar Medium. D.~M.\ thanks M.\ Shara and A.\ Pagnotta for
helpful discussions.


\end{document}